\documentclass[prl,aps,twocolumn,10pt,showpacs,floatfix,longbibliography]{revtex4-2}
\usepackage{amsfonts,amssymb,amsmath,amsthm}
\usepackage{bm}
\usepackage{braket}
\usepackage{booktabs}
\usepackage{color}
\usepackage{dcolumn}
\usepackage{dsfont}
\usepackage{float}
\usepackage{graphicx}
\usepackage{hyperref}
\usepackage{lineno}
\usepackage{mathrsfs}
\usepackage{sidecap}
\usepackage{soul}
\usepackage{subfigure}
\usepackage{verbatim}
\allowdisplaybreaks[1] 

\hypersetup{colorlinks,
linkcolor=blue,  citecolor=blue,  filecolor=blue,  urlcolor=blue }

\begin{document}	
\title{Fractional Chern Insulators in Twisted Bilayer Optical Lattices}
\author{Yan-Bin Yang$^{1}$}
\email{yangyb@tsinghua.edu.cn}
\author{Jiong-Hao Wang$^{1,2}$}
\author{Shicheng Ma$^{1}$}
\author{Yong Xu$^{1,3}$}
\email{yongxuphy@tsinghua.edu.cn}
\affiliation{$^{1}$Center for Quantum Information, IIIS, Tsinghua University, Beijing 100084, People's Republic of China}
\affiliation{$^{2}$Department of Physics, Stockholm University, AlbaNova University Center, 106 91 Stockholm, Sweden}
\affiliation{$^{3}$Hefei National Laboratory, Hefei 230088, People's Republic of China}


\begin{abstract}
Twisted bilayer materials provide a versatile platform for realizing correlated topological states.
Motivated by the recent experimental realization of atomic Bose-Einstein condensates in twisted bilayer optical lattices,
we investigate the emergence of fractional topological phases in such systems.
For experimentally realistic parameters, the single-particle spectrum hosts 
a pair of quasi-degenerate nearly flat moir\'{e} bands. 
We show using self-consistent Hartree-Fock calculations that atomic interactions 
spontaneously break time-reversal symmetry, lift the quasi-degeneracy, and 
generate an isolated nearly flat band with a nonzero Chern number.
At fractional filling, exact diagonalization of the Hamiltonian projected to this band reveals a 
threefold degenerate ground-state manifold 
and a characteristic particle entanglement spectrum, providing strong evidence for 
a fractional Chern insulator. The resulting many-body gap is of order \(10^{-3}E_R\), corresponding to 
a nanokelvin energy scale accessible to state-of-the-art cold-atom experiments. Our scheme requires 
neither spin-orbit coupling, as in transition metal dichalcogenides, 
nor a finite magnetic field, as in existing twisted-bilayer-graphene experiments, 
providing a highly tunable route to strongly correlated topological phases in twisted bilayer optical lattices.
\end{abstract}

\maketitle

Moir\'{e} superlattices in twisted bilayer materials have become
a platform for engineering a wide range of correlated quantum phases~\cite{Cao2018Nature1,Cao2018Nature2,Yankowitz2019Science,Lu2019Nature,Andrei2020NatMater,Balents2020NatPhys}.
In particular, fractional Chern insulators that exhibit the fractional quantum anomalous Hall effect at zero magnetic field~\cite{Tang2011PRL,Nupert2011PRL,Sun2011PRL,Sheng2011NC,Bernevig2011PRX} 
have been predicted and observed in twisted bilayer transition-metal dichalcogenides (TMDs)~\cite{Li2021PRResearch,Cai2023Nature,Zeng2023Nature,Park2023Nature,Xu2023PRX,Park2026NatPhys,Crepel2023PRB,Reddy2023PRB1,Reddy2023PRB2,
	WangChong2024PRL,Yu2024PRB,lu2024fractional}.
In these systems, interactions and nontrivial moir\'{e}-band topology give rise to fractional quantum anomalous Hall states.
Fractional Chern insulators have also been predicted in twisted bilayer 
graphene~\cite{Abouelkomsan2020PRL,Repellin2020PRR,Ledwith2020PRR,Wilhelm2021PRB}, 
while their experimental observation to date has involved a finite magnetic field~\cite{Xie2021Nature}.

Ultracold atoms in optical lattices provide another versatile platform for simulating
quantum many-body systems, offering precise control over lattice geometries, interactions, and detection~\cite{Gross2017Quantum,Schafer2020Tools}. 
In the context of twisted bilayer physics, it has been proposed to encode the layer degree of freedom
in atomic spin states, and to implement spin-dependent optical lattices
with a relative twist, thereby creating synthetic twisted bilayer structures~\cite{Cirac2019PRA,ZhangChuanwei2021PRL}.
Based on these proposals, twisted bilayer square lattices have recently been realized
in cold atomic gases, where a superfluid-to-Mott-insulator transition has been observed~\cite{ZhangJing2023Nature}.
Theoretical studies have also revealed a variety of interesting phenomena in twisted optical lattices~\cite{ZhangChuanwei2021PRL,GaoChao2024PRL,ZhangJing2025PRA,ZhuQizhong2025PRA1,ZhuQizhong2025PRA2,ZhuQizhong2025FrontPhys,DengYoujin2025PRB,Tian2025PRA,ZengJianhua2025QuantumFrontiers,GaoChao2026PRL},
including Larkin-Ovchinnikov superfluids~\cite{ZhangChuanwei2021PRL}, interaction-induced moir\'{e} lattices~\cite{ZhuQizhong2025PRA1}, and 
fractal band structures~\cite{GaoChao2026PRL}.
However, a strongly correlated fractional topological phase has not yet been established in twisted bilayer optical lattices. 
In particular, whether this platform can support a fractional Chern insulator remains an open question.
We note that, although various cold-atom schemes for fractional 
quantum Hall physics have been proposed outside the context of bilayer systems~\cite{Cooper2008AP,Sorensen2005PRL,Hafezi2007PRA,Cooper2013PRL,Yao2013PRL,Raciunas2016PRA,Barkeshli2015,He2017PRB,Motruk2017PRB,Hudomal2019PRA,Wang2024PRL,Palm2024PRR},
existing experimental 
realizations rely on a net magnetic flux~\cite{Greiner2023Nature}.

Here, we study moir\'{e} bands and topological phases in a twisted bilayer honeycomb optical lattice, reminiscent of a spinless twisted bilayer graphene model~\cite{Andrei2020NatMater}.
In contrast to previous proposals that rely on
exact tune-out wavelengths of lasers to realize strictly spin-selective optical potentials~\cite{Cirac2019PRA},
we choose lattice wavelengths to slightly deviate from the tune-out value.
This introduces a small additional potential for
each spin state originating from the lattice primarily designed for trapping the other spin state.
Under this setting, the noninteracting Hamiltonian exhibits nearly flat moir\'{e} bands. 
Our self-consistent Hartree-Fock (SCHF) calculations reveal that short-range atomic interactions can
lift the quasi-degeneracy of the two flat bands, giving rise to an isolated band with a nonzero Chern number.
We consider three types of HF ans\"{a}tze and find that the one assuming uniform
partial occupation of the mean-field band yields the lowest many-body ground-state energy in the parameter
regime that supports a fractional Chern insulator phase,
identified by
the ground-state degeneracy, uniform occupation in momentum space, and particle entanglement spectrum
in agreement with the generalized Pauli principle.
The many-body energy gap is on the order of $10^{-3} E_R$,
corresponding to temperatures in the nanokelvin
regime and is thus well within the reach of state-of-the-art cold-atom experimental techniques~\cite{Greiner2017Nature,Pan2020Science}.
Our setup thus provides an experimentally accessible route to realize 
fractional Chern insulators in a twisted-bilayer-graphene-like model 
under zero magnetic field, without spin-orbit coupling as in TMDs~\cite{WuFengcheng2019PRL}.

\begin{figure*}[t]
\centering
\includegraphics[width=1.0\linewidth]{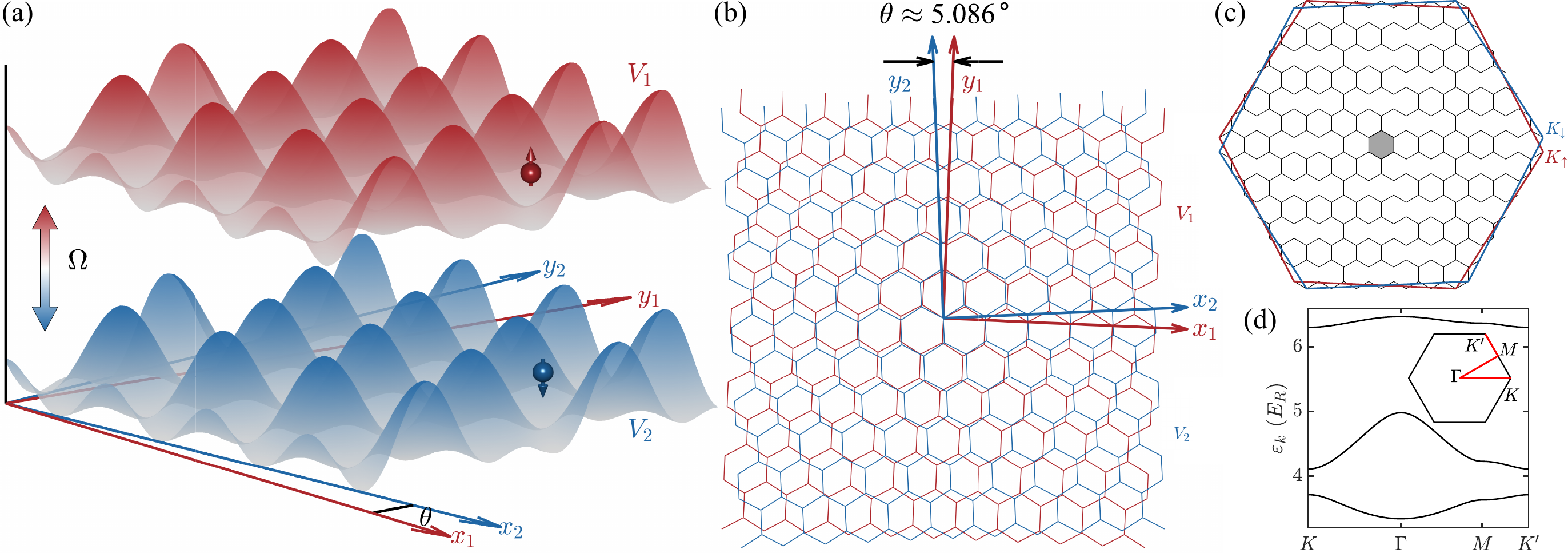}
\caption{Schematic of a twisted bilayer optical lattice formed by two spin-dependent honeycomb lattices with 
	relative twist angle $\theta$.
(a) Optical potential profiles $V_1({\bm r})$ and $V_2({\bm r})$ felt by two spin states $\ket{\uparrow}$, $\ket{\downarrow}$ simulating the two layers.
The two spin states are coupled by a microwave field with the Rabi frequency $\Omega$ simulating interlayer tunneling.
(b) Real-space moir\'{e} lattice with the rotation axis passing through a sublattice site.
(c) Moir\'{e} reciprocal lattice. The grey region marks a moir\'{e} BZ.
$K_{\uparrow}$ and $K_{\downarrow}$ denote high-symmetry momenta of 
the individual-layer BZs.
Here, we set the twist angle to the commensurate value $\theta=\arccos(253/254)\approx 5.086^\circ$.
(d) Low-energy band structure of the monolayer honeycomb optical potential $V_L({\bm r})$.
}
\label{fig1}
\end{figure*}

\emph{Single-particle Hamiltonian}---We start by considering a twisted bilayer optical
lattice implemented using spin-dependent optical potentials~\cite{Cirac2019PRA,ZhangChuanwei2021PRL},
as recently realized in an experiment with Bose-Einstein condensates~\cite{ZhangJing2023Nature}.
In the basis of the two spin states ($\ket{\uparrow}, \ket{\downarrow}$), the single-particle Hamiltonian
is given by
\begin{equation}\label{H_bilayer}
	{\hat{H}_0} =
	\left[
	\begin{array}{cc}
		\frac{\hat{\bm{p}}^2}{2 m} + V_1({\bm r}) + w V_2({\bm r}) & \Omega \\
		\Omega & \frac{\hat{\bm{p}}^2}{2 m} + V_2({\bm r}) + w V_1({\bm r}) \\
	\end{array}
	\right],
\end{equation}
where $\hat{\bm{p}}$ is the momentum operator and $m$ is the atomic mass.
The two atomic spin states encode the layer degree of freedom,
and a microwave field couples them to simulate interlayer tunneling with coupling 
strength $\Omega$.
The two spin states (layers) mainly experience the optical potential $V_1({\bm r})$ and
$V_2({\bm r})$, respectively. These potentials are obtained
by rotating an optical lattice $V_L({\bm r})$ through angles $-\theta/2$ and $\theta/2$, respectively,
where $\theta$ is the twist angle [Fig.~\ref{fig1}(a)].
The potential $V_L({\bm r})$ realizes a honeycomb lattice with broken sublattice degeneracy (see End Matter), 
opening a gap at the Brillouin zone (BZ) corners [Fig.~\ref{fig1}(d)].
The resulting twisted bilayer lattice therefore lacks the $\mathcal{C}_2 \mathcal{T}$ symmetry ($\mathcal{C}_2$ is a twofold rotation),
protecting the fragile topology in twisted bilayer graphene~\cite{BohmJung2019PRX,Song2019PRL}.
Nevertheless, the Hamiltonian preserves spinless time-reversal symmetry,
$\hat{\mathcal{T}}=\mathcal{K}$, where $\mathcal{K}$ denotes complex conjugation.
In contrast to previous studies that use tune-out wavelengths to make each
optical potential act exclusively on one spin state, corresponding to $w=0$~\cite{Cirac2019PRA,ZhangJing2023Nature,ZhangChuanwei2021PRL},
we choose wavelengths slightly detuned from the tune-out values.
Each spin state consequently experiences a weak additional potential from
the lattice intended for the other spin state, with a relative strength $w$.
We consider a commensurate twist angle,
for which the twisted bilayer lattice retains the translational symmetry~\cite{LopesdosSantos2012PRB}.
Further details of the single-particle Hamiltonian are provided in End Matter.
In the following, we take $\hbar k_L$ as the unit of
momentum and the recoil energy $E_R = \hbar^2 k_L^2 / (2 m)$ as the unit of energy,
where $k_L$ is the wavevector of lasers.

\begin{figure*}[t]
\centering
\includegraphics[width=1.0\linewidth]{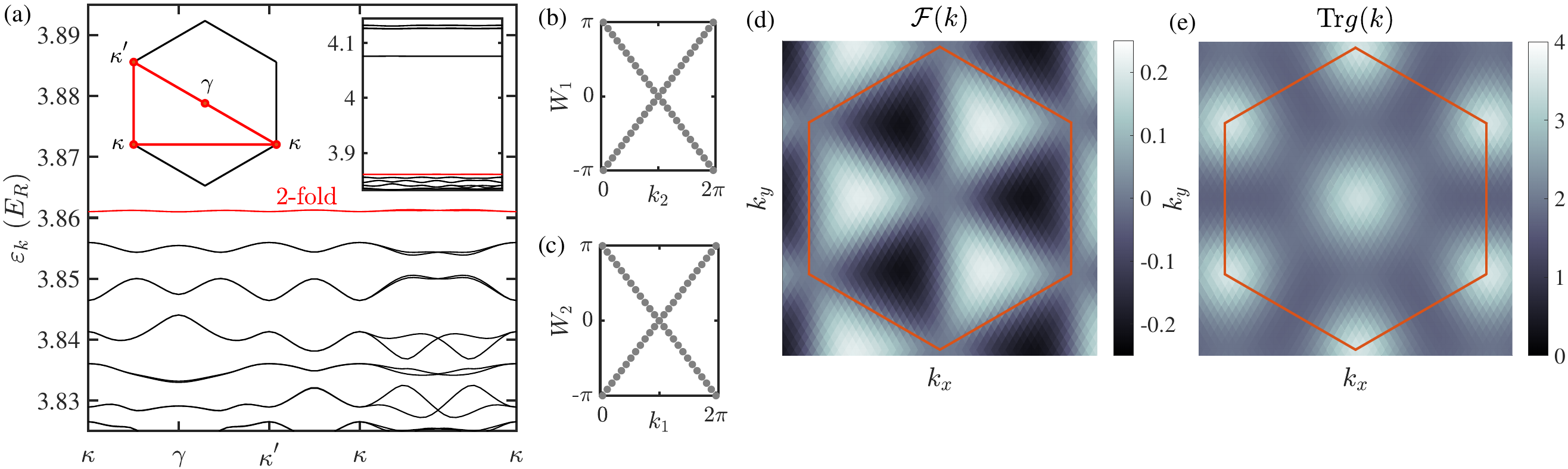}
\caption{Single-particle band properties.
(a) Moir\'{e} band structure along high-symmetry lines in the moir\'{e} BZ.
The inset shows wider energy range.
Red lines highlight the two quasi-degenerate, nearly flat bands below a large gap.
(b),(c) Wilson loop spectra of the two-band subspace along the two reciprocal basis vectors.
(d),(e) Berry curvature $\mathcal{F}({\bm k})$ and quantum metric trace $\mathrm{Tr} g({\bm k}) $ of the flat-band subspace over the moir\'{e} BZ.
Both quantities are multiplied by $S_{\text{BZ}}/(2\pi)$, where $S_{\text{BZ}}$ is the area of the moir\'{e} BZ.
}
\label{fig:SP}
\end{figure*}

We first solve the single-particle Hamiltonian to obtain the moir\'{e} band structure.
Figure~\ref{fig:SP}(a) shows the low-energy bands along high-symmetry lines in the moir\'{e} BZ.
Two quasi-degenerate, nearly flat bands lie below a gap of approximately $0.2 E_R$
---a gap much larger than all other energy scales---thus allowing us to safely neglect the higher bands in the subsequent many-body calculations.
The Wilson loop spectra of the two flat bands [Figs.~\ref{fig:SP}(b) and \ref{fig:SP}(c)] exhibit winding,
reminiscent of that in a two-dimensional time-reversal-invariant topological insulator~\cite{Alexandradinata2014PRB}. 
Here, however, spinless time-reversal symmetry does not protect this winding, and the system
is intrinsically topologically trivial.
We also examine the Berry curvature $\mathcal{F}(\bm{k})$ and the trace of quantum metric $\text{Tr}g(\bm{k}) $ 
for the two-band subspace,
because these geometric properties are closely related to the formation of fractional quantum Hall states~\cite{Roy2014PRB}.
As shown in Fig.~\ref{fig:SP}(d), the Berry curvature changes sign across the BZ and integrates to zero,
as required by time-reversal symmetry.
In contrast, the trace of quantum metric of the two bands is relatively uniform [Fig.~\ref{fig:SP}(e)]
with an average value about $2.2$,
close to the sum of contributions from two lowest Landau levels~\cite{Ozawa2021PRB}.
Although the bands are intrinsically topologically trivial,
their Wilson loop winding and the quantum metric suggest a geometric resemblance to a pair of Chern bands with Chern numbers $\pm1$.

\emph{Interaction effect}---We now study many-body phases of fermionic atoms
in the twisted bilayer optical lattice with $s$-wave contact interactions tunable through
Feshbach resonances~\cite{Chin2010RMP}. The interaction Hamiltonian is
\begin{equation}
	\hat{H}_{\text{int}} = U_{0} \int d\bm{r} \hat{n}_{\uparrow}(\bm{r}) \hat{n}_{\downarrow}(\bm{r}),
\end{equation}
where $\hat{n}_{\sigma}({\bm r})=\hat{c}^{\dagger}_{\sigma}({\bm r}) \hat{c}_{\sigma}({\bm r})$ is the 
density operator for spin $\sigma=\uparrow,\downarrow$.
This interspin interaction corresponds to an interlayer interaction in the synthetic bilayer, giving
an interaction structure distinct from that of solid-state bilayer materials.
More generally, we consider the momentum-space two-body interaction
\begin{equation}
\hat{H}_{\text{int}} = \frac{1}{2 A} \sum_{{\bm k},{\bm k}',\bm{q},\sigma,\sigma'} V_{\bm q} \,
\hat{c}^{\dagger}_{{\bm k}+{\bm q},\sigma} \hat{c}^{\dagger}_{{\bm k}'-{\bm q},\sigma'} \hat{c}_{{\bm k}',\sigma'} \hat{c}_{{\bm k},\sigma},
\end{equation}
where $\hat{c}_{{\bm k},\sigma}$ ($\hat{c}_{{\bm k},\sigma}^\dagger$) annihilates (creates) 
a particle with momentum
${\bm k}$ and spin $\sigma$, $V_{\bm q}$ is the momentum-space interaction amplitude, 
and $A$ is the system area.
For a contact interaction, $V_{\bm q}=U_0$.
We characterize the interaction strength by the dimensionless parameter 
$U:=U_{0}/(A_m E_R)$, where $A_m$ is the area of a moir\'{e} unit cell.

The single-particle spectrum contains a relatively large gap above the quasi-degenerate flat bands.
We refer to the bands above this gap as conduction bands and those below it as valence bands.
To investigate many-body topological phases at partial filling of the flat bands,
we project the full Hamiltonian onto a finite set of valence bands.
Since the flat bands lie at the top of this retained subspace, it is convenient to use a hole representation,
which regards the many-body state with fully occupied valence bands as a vacuum~\cite{Yu2024PRB}.
Defining the hole operator $b_{n,{\bm k}}:=c^{\dagger}_{n,{\bm k}}$ ($c^{\dagger}_{n,{\bm k}}$ denotes the creation operator of the Bloch state $\ket{n,{\bm k}}$),
we consider the projected Hamiltonian in the hole basis~\cite{WuFengcheng2023PRX}
\begin{widetext}
	\begin{equation}
		\hat{H}^h 
		= \hat{H}_0^h + \hat{H}_{\text{int}}^h 
		= -\sum_{n,\bm{k}} \varepsilon_{n,\bm{k}} b^{\dagger}_{n,\bm{k}} b_{n,\bm{k}}
		+\sum_{n_1,\cdots,n_4} 
		\sum_{\bm{k},\bm{k}'\in \text{BZ},\bm{q}}
		V^{n_1 n_2 n_3 n_4}_{\bm{k},\bm{k}',\bm{q}}
		b^{\dagger}_{n_1,[\bm{k}+\bm{q}]} b^{\dagger}_{n_2,[\bm{k}'-\bm{q}]} b_{n_3,\bm{k}'} b_{n_4,\bm{k}},
	\end{equation}
\end{widetext}
where $\varepsilon_{n, \bm{k}}$ is the band dispersion of the single-particle Hamiltonian $\hat{H}_0$, and
$[\cdot]$ denotes folding into the first BZ.
The projected interaction matrix element $V^{n_1 n_2 n_3 n_4}_{\bm{k},\bm{k}',\bm{q}}$ is given by
\begin{equation}\label{Eq:com}
	V^{n_1 n_2 n_3 n_4}_{\bm{k},\bm{k}',\bm{q}}
	=\left[\frac{1}{2 A} V_{\bm q} \,
	\Lambda^{n_1,n_4}(\bm{k},\bm{q})
	\Lambda^{n_2,n_3}(\bm{k}',-\bm{q})\right]^{*},
\end{equation}
where the form factors are
$\Lambda^{n_1,n_4}(\bm{k},\bm{q})=\braket{u_{n_1,\bm{k}+\bm{q}} | u_{n_4,{\bm k}}}$,
with $\ket{u_{n,\bm{k}}}$ denoting the cell-periodic part of the Bloch state $\ket{n,\bm{k}}$~\cite{SM}. 
The complex conjugation in Eq.~(\ref{Eq:com}) arises from the particle-hole transformation.

\begin{figure}[t]
	\centering
	\includegraphics[width=1.0\linewidth]{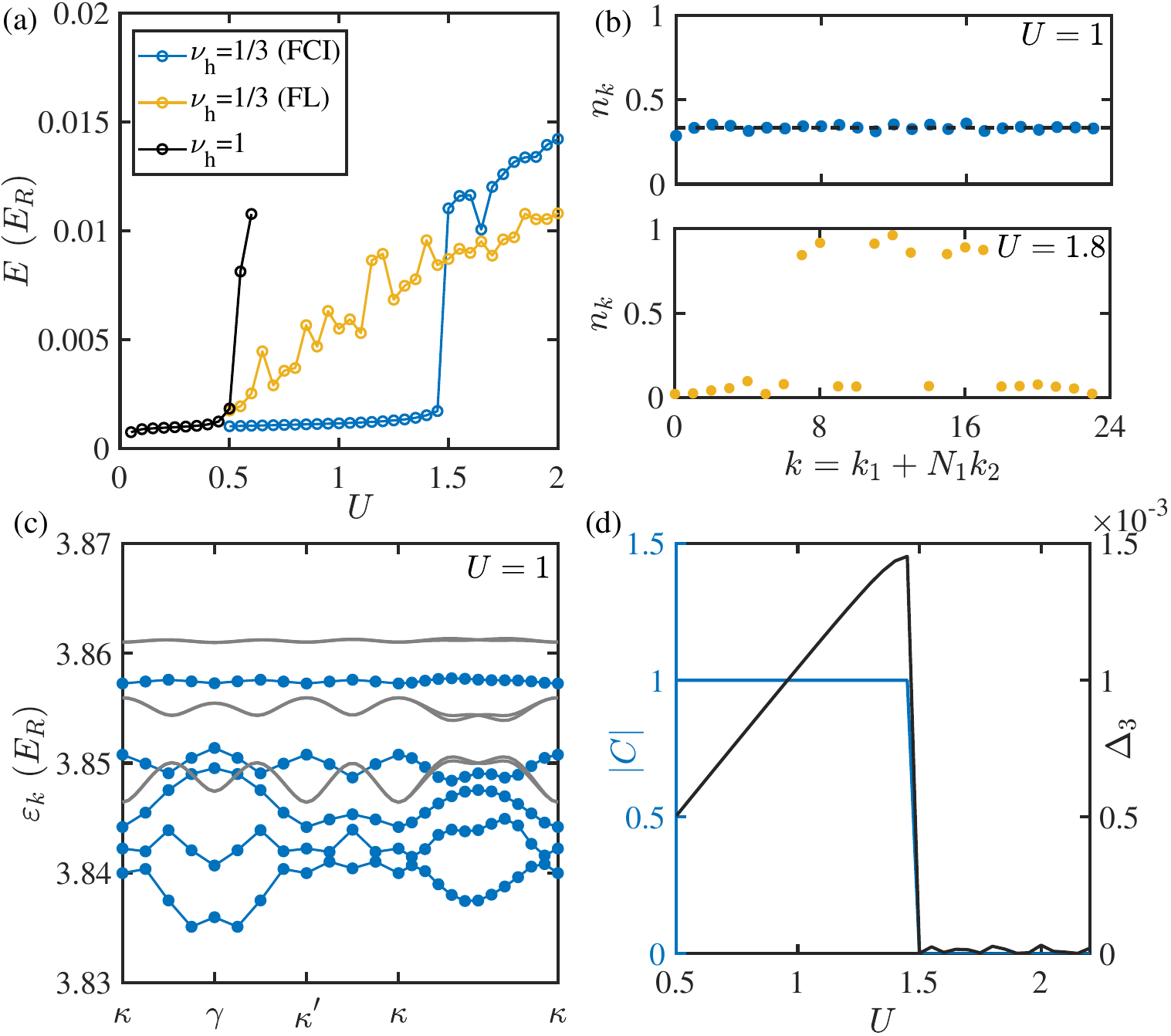}
	\caption{
		(a) ED ground-state energy with a constant subtracted for clarity, as a function of
		interaction strength $U$ for the fractional-Chern-insulator (FCI) ansatz, Fermi-liquid (FL) ansatz,
		and integer filling HF ans\"{a}tze.
		(b) Momentum-space occupation of the mean-field band in the ED ground state 
		obtained from the lowest-energy ansatz.
		(c) Mean-field band structure along high-symmetry lines obtained by
		SCHF at $U=1$ using the FCI ansatz.
		Grey lines show the noninteracting dispersion for comparison.
		(d) Absolute Chern number of the isolated mean-field band (blue line) and
		many-body gap $\Delta_3=E_4-E_3$ from ED (black line), as function of $U$.
		ED is performed for system size $(N_1, N_2) = (4, 6)$.
	}
	\label{fig:HF}
\end{figure}

\begin{figure}[t]
	\centering
	\includegraphics[width=0.95\linewidth]{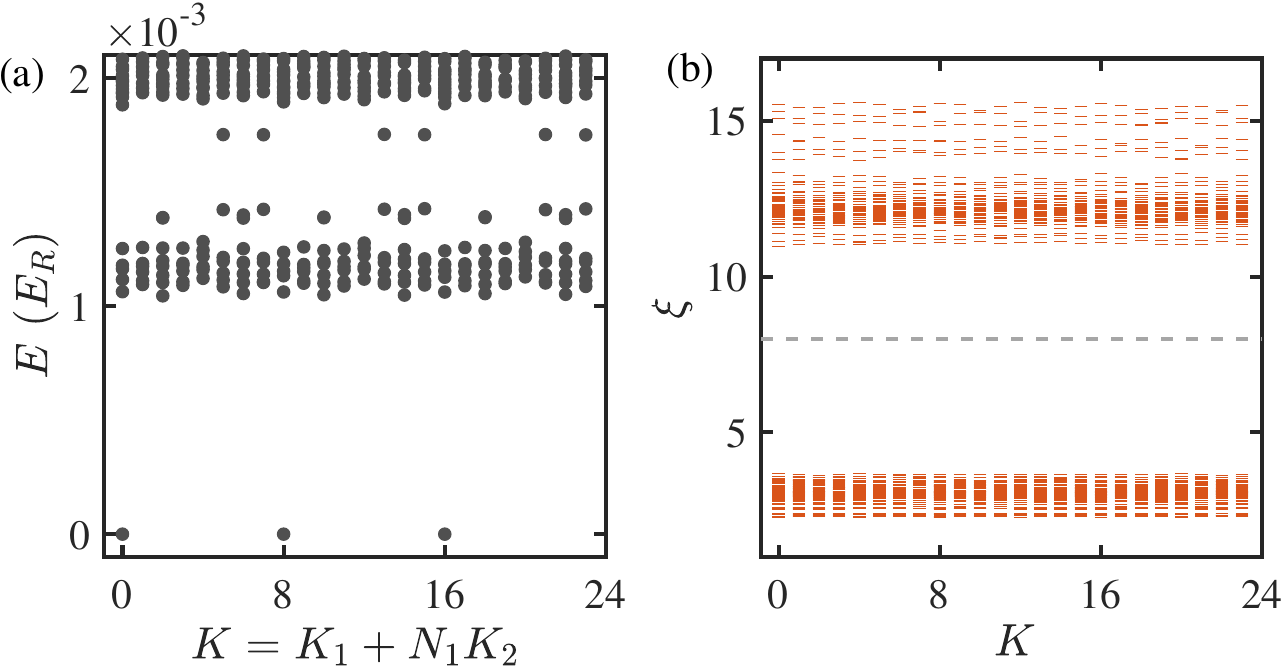}
	\caption{
		(a) Many-body energy spectrum from ED at $U=1$.
		The ground-state manifold has three-fold degeneracy with the total momenta (indexed by $K$) predicted by the generalized Pauli principle.
		(b) PES of the ground states of $N=8$ with a subsystem of $N_{A}=3$ particles.
		There are $1088$ states below an entanglement gap denoted by the dashed line, in agreement with the counting of $(1,3)$ generalized Pauli principle.
		ED is performed for system size $(N_1, N_2) = (4, 6)$.
	}
	\label{fig:ED}
\end{figure}

We consider a hole filling $\nu_h=1/3$,
equivalently to an atomic filling $\nu=5/3$ of the two flat-band subspace.
We first perform SCHF calculations to obtain interaction-renormalized bands,
and then solve the many-body Hamiltonian projected onto a mean-field band using exact diagonalization (ED).
Because the flat bands are fractionally filled, the conventional HF method at integer hole filling $\nu_h=1$ 
may overestimate the interaction effect.
A treatment adapted to fractional filling is therefore needed.
Our combined HF and ED scheme should be viewed as 
a variational approach: the HF step generates the mean-field bands as the variational 
Hilbert space for the subsequent ED calculation. The HF method with lower ground-state energy from ED is considered as better approach to construct the basis for ED calculation.

Recently, an improved HF method applicable for fractionally filled band has been proposed in studies of 
fractional quantum anomalous Hall effect in rhombohedral multilayer graphene~\cite{Huang2024PRB}.
Following this approach, we take three different HF methods to obtain the corresponding mean-field band structures
and then compare the ground-state energy obtained from ED calculations based on these different mean-field bands.
The three HF methods assume different self-consistent conditions as follows:
(i) the fractional-Chern-insulator ansatz assumes uniform partial occupation of the mean-field band over the BZ;
(ii) the Fermi-liquid ansatz assumes the occupation of the lowest-energy mean-field states at the given filling factor; and
(iii) the integer-filling ansatz assumes a fully occupied mean-field band~\cite{SM}.

Figure~\ref{fig:HF}(a) compares the ED ground-state energies obtained using
the three HF ans\"{a}tze.
We clearly see that for $0.5 < U < 1.5$, the fractional-Chern-insulator
ansatz yields the lowest energy. 
The self-consistency of the HF ans\"{a}tze are demonstrated by
the expectation values of momentum space occupation of the mean-field band
at two different interaction strengths for fractional-Chern-insulator ansatz
and Fermi-liquid ansatz, as shown in Fig.~\ref{fig:HF}(b).
Specifically, for $U=1$, the density in
momentum space calculated by ED are relatively uniform
consistent with the fractional-Chern-insulator ansatz,
whereas for $U=1.8$, the occupation exhibits a discontinuity of
Fermi surface with $1/3$ of mean field states occupied,
consistent with the Fermi-liquid ansatz.

We also see in Fig.~\ref{fig:HF}(c) that in the region of $0.5 < U < 1.5$,
the interaction opens a gap between the two bare flat bands
leaving a topmost mean-field band isolated.
Remarkably, in this region, the isolated band hosts nonzero
Chern number $|C|=1$ as shown in Fig.~\ref{fig:HF}(d),
indicating a spontaneous time-reversal symmetry breaking, even though
the noninteracting single-particle bands are topologically trivial.
As $U$ increases beyond approximately $1.5$, the mean-field band undergoes
a transition to a trivial band due to interaction-induced band mixing.
Note that, for the integer-filling ansatz, the mean-field band becomes trivial already
at $U\approx 0.5$,
and the ED ground-state energy rises rapidly,
indicating that the integer-filling ansatz of $\nu_h=1$ overestimates the interaction effect
about 3 times of the fractional-Chern-insulator ansatz at $\nu_h=1/3$.

\emph{Fractional Chern insulator}---In the ED spectrum under the fractional-Chern-insulator ansatz, we observe that when $0.5 < U < 1.5$, the system has triply degenerate ground states with
a finite many-body gap $\Delta_3=E_4-E_3$ as shown in Fig.~\ref{fig:HF}(d) and Fig.~\ref{fig:ED}(a),
strongly indicating a fractional Chern insulator phase.
In Fig.~\ref{fig:ED}(a), we also see that
the total momenta of the degenerate ground states are in exact agreement with the predictions of
the generalized Pauli principle~\cite{Bernevig2011PRX},
providing compelling evidence for a fractional Chern insulator.
Moreover, as indicated in Fig.~\ref{fig:HF}(d), the many-body gap above the ground states closes precisely when the
mean-field band becomes topologically trivial,
confirming that this fractional Chern insulator phase
arises from the fractional filling of an interaction-induced Chern band.
To further substantiate this identification, we
compute the particle entanglement spectrum (PES)~\cite{Bernevig2011PRX}.
It is obtained by partitioning the system of $N$ particles
(which, in our calculation, correspond to holes of the original Hamiltonian) 
into two subsystems consisting of $N_A$ and $N_B = N-N_A$
particles,
and then calculating the eigenvalues $\{\lambda_n = e^{-\xi_n}\}$ of the reduced density matrix
$
	\rho_A = \text{Tr}_B \frac{1}{N_d} \sum_{i=1}^{N_d} \ket{\Psi_i}\bra{\Psi_i},
$
where $\ket{\Psi_i}$ is the $i$-th many-body ground state and $N_d$ is the degeneracy of ground states.
As shown in Fig.~\ref{fig:ED}(b), the PES of $N_A=3$ in $N_1 \times N_2 = 24$ orbitals have $1088$ levels below the entanglement
gap, following the $(1,3)$-permissible counting for fractional Chern insulators~\cite{Bernevig2011PRX}.

In summary, we have demonstrated the emergence of a fractional Chern insulator phase
in an experimentally accessible twisted bilayer optical lattice.
The many-body energy gap is of order $10^{-3} E_R$,
corresponding to temperatures in the nanokelvin
regime that is accessible to state-of-the-art cold-atom experiments~\cite{Greiner2017Nature,Pan2020Science}.
In cold-atom experiments, fractional Chern insulators can be characterized via fractional Hall responses~\cite{Goldman2019PRL,Greiner2023Nature}.
Additionally, a local potential can create quasiholes or quasiparticles of fractionalized charge detectable through site-resolved imaging~\cite{LiuZhao2015PRB,DongXiaoYu2022SciPostPhys}. 
Compared with the existing bilayer moir\'{e}-material realizations, our scheme requires neither the strong spin-orbit coupling underlying topological moir\'{e} bands in TMDs nor the finite magnetic field used in the experimental realization of fractional Chern insulators in twisted bilayer graphene.
Moreover, the high tunability of twisted optical lattices provides a promising route toward realizing and characterizing interaction-driven transitions of fractional Chern insulators.
In the future, it would be interesting to extend our scheme to different types of twisted optical lattices with bosonic atoms and long-range dipolar interactions, which may give rise to exotic topological phases that are difficult to realize in solid-state moir\'{e} systems.

\begin{acknowledgments}
We thank X. Li, Q. Zhu, K. Huang, F. Wu, E. J. Bergholtz, Y. Zhang, and F. Yang for helpful discussions.
This work is supported by the National Natural Science Foundation of China (Grant No. 12474265)
and Quantum Science and Technology-National Science and Technology Major Project (Grant No. 2021ZD0301604).
Y.-B. Yang acknowledges support from the Shuimu Tsinghua Scholar Program and the Postdoctoral Fellowship Program of China Postdoctoral Science Foundation (Grant No. GZC20240878).
\end{acknowledgments}

%


\onecolumngrid

\begin{center}
\large{\textbf{End Matter}}
\end{center}

\twocolumngrid
\appendix
\renewcommand{\theequation}{A\arabic{equation}}
\setcounter{equation}{0}

\emph{Twisted bilayer honeycomb optical lattice}---In the twisted bilayer honeycomb Hamiltonian (\ref{H_bilayer}),
the two spin states ($\ket{\uparrow}, \ket{\downarrow}$) experience the optical potential 
$V_{\uparrow}({\bm r})=V_{1}({\bm r})+wV_{2}({\bm r})$ and $V_{\downarrow}({\bm r})=V_{2}({\bm r})+wV_{1}({\bm r})$, respectively.
The parameter $w$ measures the relative strength of the weak additional potential resulting from deviation from the tune-out condition.
The lattices $V_1({\bm r})$ and $V_2({\bm r})$ are obtained by
rotating a honeycomb lattice $V_L({\bm r})$ through $-\theta/2$ and $\theta/2$, respectively:
$V_{1,2}({\bm r}) = V_L(R_{\pm \theta/2}{\bm r})$,
where $R_{\alpha}$ denotes a rotation by an angle $\alpha$.
$\theta$ is the twist angle.
A honeycomb optical lattice is generated by interfering three laser beams separated by $120^{\circ}$ in the $x-y$ plane
tightly confined by additional lasers in the $z$ direction~\cite{Li2016Science}.
The optical potential $V_L(\bm{r})$ is given by the AC Stark shift as
\begin{equation*}
	V_L(\bm{r}) = V_0 \left|\sum_{j=1}^{3} \bm{\epsilon}_j \exp[i\bm{k}_j \cdot (\bm{r}-\bm{r}_0)]\right|^2,
\end{equation*}
where $V_0$ sets the lattice depth, $\bm{k}_j$ and $\bm{\epsilon}_j$ are
the wavevectors and the polarization vectors of the three beams, respectively, and the constant vector
${\bm r}_0$ specifies the lattice origin.
The wavevectors are $\bm{k}_1=k_L[1,0,0]^\top$, $\bm{k}_2=k_L[-1/2,\sqrt{3}/2,0]^\top$, and $\bm{k}_3=k_L[-1/2,-\sqrt{3}/2,0]^\top$,
The polarization vectors are $\bm{\epsilon}_{j}=\cos(\theta_p)\bm{\epsilon}^s + \sin(\theta_p) \bm{\epsilon}^p_j$,
where $\theta_p$ controls the mixing of the out-of-plane ($s$) and in-plane ($p$) components.
The out-of-plane component is $\bm{\epsilon}^s=[0,0,1]^\top$, and the in-plane components are
$\bm{\epsilon}^p_1=[0,1,0]^\top$, $\bm{\epsilon}^p_2=e^{i2\pi/3}[-\sqrt{3}/2,-1/2,0]^\top$, and $\bm{\epsilon}^p_3=e^{i4\pi/3}[\sqrt{3}/2,-1/2,0]^\top$.
The three beams generate an optical potential with minima on the $A$ and $B$ sublattices of a honeycomb lattice.
A nonzero $\theta_p$ lifts the $A$-$B$ sublattice degeneracy and opens a gap at the Dirac points $K$ and $K'$.

The main-text calculations use $V_0=2.0$, $\theta_p=0.09\pi$,
$\Omega=0.14 E_R$, and $w=0.014$, with the commensurate twist angle 
$\theta=\arccos(253/254)\approx 5.086^\circ$.
For honeycomb or triangular lattices,
a commensurate twist angle can be generated by two coprime integers $(m, n)$~\cite{LopesdosSantos2012PRB}:
\begin{equation*}
	\theta = \arccos \frac{3 m^2 + 3 m n + n^2/2}{3 m^2 + 3 m n + n^2}.
\end{equation*}
We choose $(m,n)=(6,1)$.

The real-space and reciprocal-space basis vectors of the monolayer honeycomb lattice are
\begin{equation*}
	\bm{a}_{1,2} = a_0\left(\pm \frac{1}{2},\frac{\sqrt{3}}{2}\right), \quad
	\bm{g}_{1,2} = \frac{4\pi}{\sqrt{3}a_0}\left(\pm \frac{\sqrt{3}}{2},\frac{1}{2}\right),
\end{equation*}
where $a_0=4\pi/(3 k_L)$.
The moir\'{e} superlattice 
has an enlarged real-space unit cell and a correspondingly reduced reciprocal-space unit cell. 
We focus on commensurate structures specified by integer pairs $(m, n)$ with $n=1$ for which
the moir\'{e} lattice vectors and reciprocal lattice vectors are
\begin{equation*}
	\bm{t}_{1,2} = a_m \left(\frac{\sqrt{3}}{2}, \mp \frac{1}{2}\right), \quad
	\bm{G}_{1,2} = \frac{4\pi}{\sqrt{3}a_m}\left(\frac{1}{2}, \mp \frac{\sqrt{3}}{2} \right),
\end{equation*}
where $a_m = a_0/(2\sin(\theta/2))$.

\emph{Wilson loop}---We calculate the Wilson loop using the discretized expression
\[
\mathcal{W}_{\mathcal C}
=
\prod_{j=1}^{N}
\mathcal{O}(\bm{k}_j,\bm{k}_{j+1}),
\]
where \(\mathcal{O}_{mn}(\bm{k}_j,\bm{k}_{j+1})
=
\langle u_m(\bm{k}_j)|u_n(\bm{k}_{j+1})\rangle \), and \(\ket{u_n(\bm{k})}\) is the cell-periodic Bloch state of the $n$-th band in the selected band subspace. The closed momentum-space path \(\mathcal C\) is discretized into \(N\) points, with \(\bm{k}_1\equiv \bm{k}_{N+1}\).
In the limit \(N \to \infty\), \(\mathcal{W}_{\mathcal{C}}\) is unitary and has eigenvalues \(\mathrm{e}^{\mathrm{i}\theta_n}\).
The phases \(\{\theta_n\}\) form the Wilson loop spectrum and encode the corresponding Wannier centers~\cite{Alexandradinata2014PRB}. 
Specifically, we evaluate loops traversing the BZ along one reciprocal basis vector and plot the resulting spectrum as a function of the crystal momentum along the other.

\emph{Quantum geometry}---The quantum geometry of Bloch states is characterized by the quantum geometric tensor (QGT) defined as~\cite{HerzogArbeitman2022PRL},
\[
\mathcal{Q}_{ij}(\bm{k})=
\text{Tr} 
\left[ \partial_{\bm{k}_i} U^\dagger(\bm{k}) \left(1 - P(\bm{k}) \right) \partial_{\bm{k}_j} U(\bm{k}) \right],
\]
where the columns of the $N_{\mathrm{orb}}\times N_{\mathrm{band}}$ matrix $U(\bm{k})$ are the Bloch states spanning the selected band subspace,
\(P(\bm{k}) = U(\bm{k}) U^\dagger(\bm{k})\) is the gauge-invariant projector onto that subspace, and
\(\bm{k}_i\) is the \(i\)-th momentum component.
In this expression, \(\text{Tr}\) denotes the trace over the selected bands.
The QGT is gauge invariant and decomposes into a symmetric real part and an antisymmetric imaginary part:
\[
\mathcal{Q}_{ij}(\bm{k}) = g_{ij}(\bm{k}) - \frac{\mathrm{i}}{2} \mathcal{F}_{ij}(\bm{k}),
\]
where
\begin{align*}
	g_{ij}(\bm{k}) &= \frac{1}{2}\text{Tr} \left(P(\bm{k}) \{\partial_{k_i}P(\bm{k}), \partial_{k_j}P(\bm{k})\}\right), \\
	\mathcal{F}_{ij}(\bm{k}) &= \mathrm{i}\text{Tr} \left(P(\bm{k})[\partial_{k_i}P(\bm{k}),\partial_{k_j}P(\bm{k})]\right).
\end{align*}
The trace of the quantum metric and the Berry curvature are defined as
\begin{align*}
	\mathrm{Tr} g({\bm k}) &:= \sum_{i=x,y} g_{ii}(\bm{k}), \\
	\mathcal{F}(\bm{k}) &:= \mathcal{F}_{xy}(\bm{k}).
\end{align*}
For the lowest Landau level, which supports Laughlin states, both quantities are uniform across the BZ: 
\(\mathrm{Tr} g({\bm k}) = |\mathcal{F}(\bm{k})| = l_{B}^2\), where \(l_B\) is the magnetic length~\cite{Ozawa2021PRB}.


\clearpage
\onecolumngrid

\begin{center}
\large{\textbf{Supplemental Material}}
\end{center}
\setcounter{equation}{0} \setcounter{figure}{0} \setcounter{table}{0} %
\renewcommand{\theequation}{S\arabic{equation}}
\renewcommand{\thefigure}{S\arabic{figure}}

\section{Single-particle bands}
In this section, we describe the calculation of single-particle states in twisted bilayer optical lattices. 
The moir\'{e} bands $ \varepsilon_{n,\bm{k}} $ and Bloch states $ \ket{\psi_{n,\bm{k}}} $ are obtained
in a plane-wave basis:
\begin{align}
	&\hat{H}_0\ket{\psi_{n,\bm{k}}} = \varepsilon_{n,\bm{k}}\ket{\psi_{n,\bm{k}}}, \\
	&\ket{\psi_{n,\bm{k}}} = \sum_{\bm{g},s} u_{n,\bm{k}}(\bm{g},s) \ket{\bm{k} + \bm{g}, s},
\end{align}
where $\bm{k}$ lies within the moir\'{e} BZ, $\bm{g}$ is a moir\'{e} reciprocal lattice vector, and $\ket{\bm{k}, s}$ denotes a plane wave state with momentum $\bm{k}$ and spin (layer) component $s$.

Alternatively, we can diagonalize the Hamiltonian in the basis of the low-energy monolayer bands~\cite{ZhangChuanwei2021PRL}.
We first calculate the Bloch states
$\ket{\phi_{m,\bm{q}}^{l=1,2}}$
of the monolayer potentials $V_{l=1,2}(\bm{r})$:
\begin{equation}
	\left[\frac{\hat{\bm{p}}^2}{2 m} + V_{l}({\bm r})\right]\ket{\phi_{m,\bm{q}}^{l}} = \epsilon_{m,\bm{q}}^{l}
	\ket{\phi_{m,\bm{q}}^{l}}.
\end{equation}
We then diagonalize $\hat{H}_0$ in the basis of these monolayer Bloch states.
In this basis,
the matrix elements of the interlayer coupling, proportional to $\Omega$, and the additional intralayer potential, proportional to $w$, can be evaluated directly. Since a large gap separates the two lowest-energy monolayer bands from higher bands, we retain the low-energy subspace with $m=1,2$
to obtain the moir\'{e} bands near the Dirac-point gap, which are relevant to the fractional Chern insulator phase studied in the main text.

\section{Interaction Hamiltonian}
In this section, we derive the interaction Hamiltonian in the band basis. 
Consider a generic two-body interaction between two-component fermions,
\begin{equation}
	\hat{H}_{\text{int}} = \frac{1}{2} \sum_{s,s'} \int d \bm{r} d \bm{r}' 
	V^{s s'}(\bm{r}-\bm{r}') \hat{n}_{s}(\bm{r}) \hat{n}_{s'}(\bm{r'}),
\end{equation}
where $\hat{n}_{s}(\bm{r}) = \hat{c}_{s}^\dagger(\bm{r}) \hat{c}_{s}(\bm{r})$ is the density operator for spin component $s$.
In the main text, we assume a spin-independent interaction kernel, $V^{s s'}(\bm{r}) = V(\bm{r})$ with no dependence on $s,s'$.
Applying the Fourier transform $\hat{c}_{s}(\bm{r}) = \frac{1}{\sqrt{A}} \int d{\bm{k}} e^{i\bm{k}\cdot\bm{r}} \hat{c}_{\bm{k},s}$,
we obtain the interaction Hamiltonian in momentum space
\begin{equation}
	\hat{H}_{\text{int}} = \frac{1}{2 A} \sum_{{\bm k},{\bm k}',\bm{q},s,s'} V^{s s'}_{\bm q} \,
	\hat{c}^{\dagger}_{{\bm k}+{\bm q},s} \hat{c}^{\dagger}_{{\bm k}'-{\bm q},s'} \hat{c}_{{\bm k}',s'} \hat{c}_{{\bm k},s},
\end{equation}
where $V^{s s'}_{\bm q} = \int_{\bm{r}} e^{- i\bm{q}\cdot\bm{r}} V^{s s'}(\bm{r})$,
and $A$ is the system area.
The interaction kernel satisfies the Hermiticity condition
$V^{s s'}(\bm{r})=[V^{s' s}(-\bm{r})]^{*}$, or equivalently  $V^{s s'}_{\bm{q}}=[V^{s' s}_{\bm q}]^{*}$. 

Diagonalizing the single-particle Hamiltonian gives
\begin{equation}
	\hat{H}_0 = \sum_{n,\bm{k}} \varepsilon_{n,\bm{k}} \hat{c}_{n,\bm{k}}^\dagger \hat{c}_{n,\bm{k}},
\end{equation}
where $\hat{c}_{n,\bm{k}}^\dagger$ creates the Bloch state $ \ket{\psi_{n,\bm{k}}} $.
The Bloch states have the plane-wave expansion
\begin{equation}
	 \ket{\psi_{n,\bm{k}}} =e^{i\bm{k}\cdot \hat{\bm{r}}}\ket{u_{n,\bm{k}}} 
	=\sum_{\bm{g},s} u_{n,\bm{k}}(\bm{g},s) \ket{\bm{k}+\bm{g},s},
\end{equation}
which yields the transformation between the plane wave states and the band states 
$\hat{c}_{\bm{k}+\bm{g},s} = \sum_{n} u_{n,\bm{k}}(\bm{g},s) \hat{c}_{n,\bm{k}}$, 
with crystal momentum $\bm{k} \in \text{BZ}$ and reciprocal lattice vector $\bm{g}$.
The interaction Hamiltonian projected onto the Bloch bands is then
\begin{equation}
	\hat{H}_{\text{int}}  
	= \sum_{n_1,\cdots,n_4} 
	\sum_{\bm{k},\bm{k}'\in \text{BZ},\bm{q}}
	V^{n_1 n_2 n_3 n_4}_{\bm{k},\bm{k}',\bm{q}}
	c^{\dagger}_{n_1,[\bm{k}+\bm{q}]} c^{\dagger}_{n_2,[\bm{k}'-\bm{q}]} 
	c_{n_3,\bm{k}'} c_{n_4,\bm{k}},
\end{equation}
where $[\bm{k}]$ denotes momentum folded into the first BZ.
The projected interaction matrix element is
\begin{equation}
	V^{n_1 n_2 n_3 n_4}_{\bm{k},\bm{k}',\bm{q}}
	=\frac{1}{2 A}\sum_{s,s'} V^{s s'}_{\bm q} \,
	\Lambda^{n_1 n_4}_{s}(\bm{k},\bm{q})
	\Lambda^{n_2 n_3}_{s'}(\bm{k}',-\bm{q}),
\end{equation}
with form factors
\begin{equation}
	\Lambda^{n_1 n_4}_s(\bm{k},\bm{q})=\braket{u_{n_1,\bm{k}+\bm{q}} | u_{n_4,{\bm k}}}_s :=
	\sum_{\bm{g}} u^{*}_{n_1,[\bm{k}+\bm{q}]}(\bm{g}+p(\bm{k}+\bm{q}),s) u_{n_4,\bm{k}}(\bm{g},s).
\end{equation}
Here, $p(\bm{k}):=\bm{k}-[\bm{k}]$ is a reciprocal lattice vector.
We adopt a periodic gauge in which $\ket{n,\bm{k}+\bm{g}}=\ket{n,\bm{k}}$ and
$u_{n,\bm{k}+\bm{g}}(\bm{g}',s)=u_{n,\bm{k}}(\bm{g}+\bm{g}',s)$.

\section{Self-consistent Hartree-Fock method}

In this section, we describe the SCHF method for the interacting Hamiltonian projected onto the noninteracting bands:
\begin{align}
	\hat{H} &= \hat{H}_0 + \hat{H}_{\text{int}}, \\
	\hat{H}_0 &= \sum_{n,\bm{k}} \varepsilon_{n,\bm{k}} \hat{c}_{n,\bm{k}}^\dagger \hat{c}_{n,\bm{k}}, \\
	\hat{H}_{\text{int}} &=
	\sum_{n_1,\cdots,n_4} 
	\sum_{\bm{k},\bm{k}'\in \text{BZ},\bm{q}}
	V^{n_1 n_2 n_3 n_4}_{\bm{k},\bm{k}',\bm{q}}
	\hat{c}^{\dagger}_{n_1,[\bm{k}+\bm{q}]} \hat{c}^{\dagger}_{n_2,[\bm{k}'-\bm{q}]} 
	\hat{c}_{n_3,\bm{k}'} 
	\hat{c}_{n_4,\bm{k}}.
\end{align}
Here, $\hat{c}_{n,\bm{k}}$ ($\hat{c}^{\dagger}_{n,\bm{k}}$) denotes the annihilation (creation) operator for the Bloch state $\ket{n,\bm{k}}$ with the band energy $\varepsilon_{n,\bm{k}}$.
The formulas here apply to the general SCHF formalism, in which $\hat{c}_{n,\bm{k}}$ and $\hat{c}^{\dagger}_{n,\bm{k}}$ denote generic annihilation and creation operators, respectively. 
In the main text, we instead work in the hole basis, with the particle operators and band dispersions replaced by their hole counterparts.

In the HF approximation, the mean-field Hamiltonian is expressed in terms of the one-body density matrix
$P_{m n}(\bm{k},\bm{k}'):=
\braket{\hat{c}^\dagger_{m,\bm{k}} \hat{c}_{n,\bm{k}'}}$.
We assume the many-body state preserves the lattice translational symmetry, so that the density matrix is diagonal in the crystal momentum: $P_{m n}(\bm{k},\bm{k}') = P_{m n}(\bm{k})\delta_{\bm{k},\bm{k}'}$.
The interaction Hamiltonian is then decoupled as
\begin{align}
	\hat{H}^{\text{HF}}_{\text{int}} &= \hat{H}_{\text{Hartree}} + \hat{H}_{\text{Fock}}, \\
	\hat{H}_{\text{Hartree}} &= \sum 
	V^{n_1 n_2 n_3 n_4}_{\bm{k},\bm{k}',\bm{q}} \left( \braket{\hat{c}^{\dagger}_{n_1,[\bm{k}+\bm{q}]}\hat{c}_{n_4,\bm{k}}}\hat{c}^{\dagger}_{n_2,[\bm{k}'-\bm{q}]} 
	\hat{c}_{n_3,\bm{k}'}
	+ \braket{\hat{c}^{\dagger}_{n_2,[\bm{k}'-\bm{q}]} 
		\hat{c}_{n_3,\bm{k}'}}\hat{c}^{\dagger}_{n_1,[\bm{k}+\bm{q}]}\hat{c}_{n_4,\bm{k}} \right. \nonumber \\
		&\quad \left. -\braket{\hat{c}^{\dagger}_{n_1,[\bm{k}+\bm{q}]}\hat{c}_{n_4,\bm{k}}}\braket{\hat{c}^{\dagger}_{n_2,[\bm{k}'-\bm{q}]} 
			\hat{c}_{n_3,\bm{k}'}}
	\right), \nonumber \\
	&= \sum V^{n_1 n_2 n_3 n_4}_{\bm{k},\bm{k}',\bm{g}}
	\left(P_{n_1 n_4}(\bm{k}) \hat{c}^{\dagger}_{n_2,\bm{k}'} 
	\hat{c}_{n_3,\bm{k}'} + 
	P_{n_2 n_3}(\bm{k}') \hat{c}^{\dagger}_{n_1,\bm{k}} 
	\hat{c}_{n_4,\bm{k}} - P_{n_1 n_4}(\bm{k})P_{n_2 n_3}(\bm{k}')\right) , \\
	\hat{H}_{\text{Fock}} &= \sum 
	V^{n_1 n_2 n_3 n_4}_{\bm{k},\bm{k}',\bm{q}} \left( -\braket{\hat{c}^{\dagger}_{n_1,[\bm{k}+\bm{q}]}\hat{c}_{n_3,\bm{k}'}}\hat{c}^{\dagger}_{n_2,[\bm{k}'-\bm{q}]} 
	\hat{c}_{n_4,\bm{k}}
	- \braket{\hat{c}^{\dagger}_{n_2,[\bm{k}'-\bm{q}]} 
		\hat{c}_{n_4,\bm{k}}}\hat{c}^{\dagger}_{n_1,[\bm{k}+\bm{q}]}\hat{c}_{n_3,\bm{k}'} \right. \nonumber \\
	&\quad \left. +\braket{\hat{c}^{\dagger}_{n_1,[\bm{k}+\bm{q}]}\hat{c}_{n_3,\bm{k}'}}\braket{\hat{c}^{\dagger}_{n_2,[\bm{k}'-\bm{q}]} 
		\hat{c}_{n_4,\bm{k}}}
	\right), \nonumber \\
	&= \sum V^{n_1 n_2 n_3 n_4}_{\bm{k},\bm{k}',\bm{g}+\bm{k}'-\bm{k}}
	\left(-P_{n_1 n_3}(\bm{k}') \hat{c}^{\dagger}_{n_2,\bm{k}} 
	\hat{c}_{n_4,\bm{k}} - 
	P_{n_2 n_4}(\bm{k}) \hat{c}^{\dagger}_{n_1,\bm{k}'} 
	\hat{c}_{n_3,\bm{k}'} + P_{n_1 n_3}(\bm{k}') P_{n_2 n_4}(\bm{k})\right) , 
\end{align}
where $\bm{k}, \bm{k}'$ denote momenta in the first $\text{BZ}$, $\bm{g}$ denotes reciprocal lattice vector, and terms beyond first order of $ \hat{c}^\dagger_{m,\bm{k}} \hat{c}_{n,\bm{k}} - P_{m n}(\bm{k}) $ are neglected.
Including the noninteracting part, the mean-field Hamiltonian takes the quadratic form
\begin{equation}
	\hat{H}^{\text{HF}} = \sum_{\bm{k}}\bigg[\sum_{m,n} \hat{c}^\dagger_{m,\bm{k}} h_{m n}^{\text{HF}}(\bm{k}) \hat{c}_{n,\bm{k}} - E_0(\bm{k})\bigg],
\end{equation}
where $E_0(\bm{k}) = \frac{1}{2}\text{Tr}\left\{[ h^\text{Hartree}(\bm{k})+h^\text{Fock}(\bm{k})]P^{\top}(\bm{k}) \right\}$.
The HF matrix is $h^{\text{HF}}_{m n}(\bm{k}) =\varepsilon_{n,\bm{k}}\delta_{m,n}+h_{m n}^{\text{Hartree}}(\bm{k}) + h_{m n}^{\text{Fock}}(\bm{k})$, with
\begin{align}
	&h^{\text{Hartree}}_{m n}(\bm{k}) = \sum_{s,s',\bm{g},m',n',\bm{k}'} \frac{1}{2A} \left(V^{s s'}_{\bm{g}}+V^{s' s}_{-\bm{g}}\right) \Lambda^{m n}_{s}(\bm{k},\bm{g})
	\Lambda^{m' n'}_{s'}(\bm{k}',-\bm{g}) P_{m' n'}(\bm{k}'), \\
	&h^{\text{Fock}}_{m n}(\bm{k}) = \sum_{s,s',\bm{g},m',n',\bm{k}'} -\frac{1}{2A} \left(V^{s s'}_{\bm{g}+\bm{k}'-\bm{k}}+V^{s' s}_{-\bm{g}-\bm{k}'+\bm{k}}\right) 
	\Lambda^{m' n}_{s}(\bm{k},\bm{g}+\bm{k}'-\bm{k})
	\Lambda^{m n'}_{s'}(\bm{k}',-\bm{g}-\bm{k}'+\bm{k}) P_{m' n'}(\bm{k}').
\end{align}
The eigenvalues of $h^{\text{HF}}(\bm{k})$ give the mean-field band structure. 
The total HF energy is given by the expectation value of $\hat{H}^{\text{HF}}$ with respect to the many-body state satisfying the self-consistency condition.
Following Ref.~\cite{Huang2024PRB}, we use three occupation constraints in the main-text SCHF calculations. 
As in the main text, we consider the lowest-energy mean-field band is partially filled with the filling $\nu$.

(i) FCI ansatz. We assume uniform occupation of the lowest-energy mean-field band at fractional filling $\nu$ throughout the BZ. 
The density matrix is given by
\begin{equation}
	P_{ab}(\bm{k}) = \nu v_a^{*} v_b,
\end{equation}
where $v_a$ is the $a$-th component of the lowest-energy eigenvector of $h^{\text{HF}}(\bm{k})$.
This constraint is motivated by the uniform orbital occupation of fractional quantum Hall states.

(ii) FL ansatz. We assume the lowest-energy mean-field states are occupied. The density matrix is given by
\begin{equation}
	P_{ab}(\bm{k}) = n_{\bm{k}} v_a^{*} v_b,
\end{equation}
where $n_{\bm{k}}$ is the occupation number of the mean-field state at $\bm{k}$, satisfying $\sum_{\bm{k}} n_{\bm{k}} = \nu N_{\text{uc}}$. 
Here, $N_{\text{uc}}$ denotes the number of unit cells.

(iii) Integer-filling ansatz. We assume that the lowest-energy mean-field band is fully occupied, giving
 \begin{equation}
 	P_{ab}(\bm{k}) = v_a^{*} v_b.
 \end{equation}

For each ansatz, we solve the mean-field Hamiltonian iteratively. At the $i$-th step, we diagonalize the quadratic Hamiltonian $h^{\text{HF}}(\bm{k})$
constructed from the density matrices $P^{(i)}(\bm{k})$.
We then construct an updated density matrix $P^{(i+1)}(\bm{k})$
using the corresponding occupation constraint and repeat the procedure until both the density matrix and the total HF energy converge. 
The converged solution of SCHF calculation yields the mean-field bands and their Bloch states. The Bloch states of the lowest-energy mean-field band are given by
\begin{equation}
	\ket{\bm{k}}^{\text{HF}} = \sum_{a} v^{\text{HF}}_a \ket{a,\bm{k}},
\end{equation}
where $\ket{a,\bm{k}}$ is a Bloch state of a noninteracting band, and
$v^{\text{HF}}_a$ is the $a$-th component of the lowest-energy eigenvector of the converged matrix $h^{\text{HF}}(\bm{k})$.
Finally, we project the many-body Hamiltonian onto the subspace spanned by $\{ \ket{\bm{k}}^{\text{HF}} \}$, and solve it by ED.
The lowest ED ground-state energy selects the optimal SCHF ansatz among the three candidates.

\end{document}